# Migrating Multi-page Web Applications to Single-page Ajax Interfaces

Ali Mesbah and Arie van Deursen

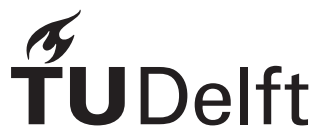
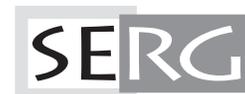

# Migrating Multi-page Web Applications to Single-page AJAX Interfaces

**Ali Mesbah**
*Software Engineering Research Group*
*Delft University of Technology*
*The Netherlands*
*A.Mesbah@tudelft.nl*

**Arie van Deursen**
*Software Engineering Research Group*
*Delft Univ. of Technology & CWI*
*The Netherlands*
*Arie.vanDeursen@tudelft.nl*

**Abstract**

*Recently, a new web development technique for creating interactive web applications, dubbed* AJAX, *has emerged. In this new model, the single-page web interface is composed of individual components which can be updated/replaced independently. If until a year ago, the concern revolved around migrating legacy systems to web-based settings, today we have a new challenge of migrating web applications to single-page* AJAX *applications. Gaining an understanding of the navigational model and user interface structure of the source application is the first step in the migration process.*

*In this paper, we explore how reverse engineering techniques can help analyze classic web applications for this purpose. Our approach, using a schema-based clustering technique, extracts a navigational model of web applications, and identifies candidate user interface components to be migrated to a single-page* AJAX *interface. Additionally, results of a case study, conducted to evaluate our tool, are presented.*

## 1. Introduction

Despite their enormous popularity, web applications have suffered from poor interactivity and responsiveness towards end users. Interaction in classic web applications is based on a multi-page interface model, in which for every request the entire interface is refreshed.

Recently, a new web development technique for creating interactive web applications, dubbed AJAX (Asynchronous JavaScript And XML) [14], has emerged. In this new model, the single-page web interface is composed of individual components which can be updated/replaced independently, so that the entire page does not need to be reloaded on each user action. This, in turn, helps to increase the levels of interactivity, responsiveness and user satisfaction [20].

Adopting AJAX-based techniques is a serious option not only for newly developed applications, but also for existing web sites if their user friendliness is inadequate. Many organizations are beginning to consider migration (ajaxification) possibilities to this new paradigm which promises rich interactivity and satisfaction for their clients. As a result, the well-known problem of legacy migration is becoming increasingly important for web applications. If until a year ago, the problem revolved around migrating legacy systems to web applications, today we have a new challenge of migrating classic web applications to single-page web applications.

The main question addressed in this paper is how to identify appropriate candidate single-page components from a page sequence interface web application. Obtaining a clear understanding of the navigational model and user interface structure of the source application is an essential step in the migration process.

In this paper, we present a reverse engineering technique for classification of web pages. We use a schema-based clustering technique to classify web pages with similar structures. These clusters are further analyzed to suggest candidate user interface components for the target AJAX application.

The rest of this paper is organized as follows. We start out, in Section 2 by exploring AJAX and focusing on its characteristics. Section 3 presents the overall picture of the migration process. Section 4 describes our page classification notion and proposes a schema-based clustering approach. Section 5 outlines how we identify candidate user interface components. The implementation details of our tool, called RETJAX, are explained in Section 6. Section 7 evaluates a sample web application and its recovered navigational and component model by applying RETJAX. Section 8 discusses the results and open issues. Section 9 covers related work. Finally, Section 10 draws conclusions and presents future work.

## 2. AJAX

As defined by Garrett [14], AJAX incorporates: standards-based presentation using XHTML and CSS, dynamic display and interaction using the Document Object Model, data interchange and manipulation, asynchronous data retrieval using XMLHttpRequest, and JavaScript binding everything together.

AJAX is an approach to web application development utilizing a combination of established web technologies to provide a more interactive web-based user interface. It is the combination of these technologies that makes AJAX unique on the Web. Well known examples of AJAX web applications include Google Suggest, Google Maps, Flickr, Gmail and the new version of Yahoo! Mail.

AJAX enables web developers to create web applications based on a single-page interface model in which the client-/server interaction is based merely on state changes. The communication between client and server can take place asynchronously, which is substantially different from the classic synchronous request, wait for response, and continue model [20].

Figure 1 shows a meta-model of a single-page AJAX web application which is composed of widgets. Each widget, in turn, consists of a set of user interface components. The specific part of the meta-model is target specific, i.e., each AJAX framework provides a specific set of UI components at different levels of granularity. The client side page is composed of client-side views, which are generated by the server-side widgets/components. Navigation is through view changes. For each view change, merely the state changes are interchanged between the client and the server, as opposed to the full-page retrieval approach in multi-page web applications.

The architectural decisions behind AJAX change the way we develop web applications. Instead of thinking in terms of sequences of Web pages, Web developers can now program their applications in the more intuitive single-page user interface (UI) component-based fashion along the lines of, for instance, Java AWT and Swing.

An overview of the architectural – processing, connecting, and data – elements of AJAX, and the constraints that should hold between them in order to meet such properties as user interactivity, scalability, and portability is given by our SPIAR [20] architectural style for AJAX applications.

## 3. Migration Process

What we would like to achieve is support in migration from a multi-page web application to a single-page AJAX interface. In this section we describe the steps needed in such a process.

Figure 2 depicts an overall view of the migration process. Note that we are primarily focusing on the user interface and not on the server-side code (which is also an essential part of a migration process). The user interface migration process consists of five major steps:

1. Retrieving Pages

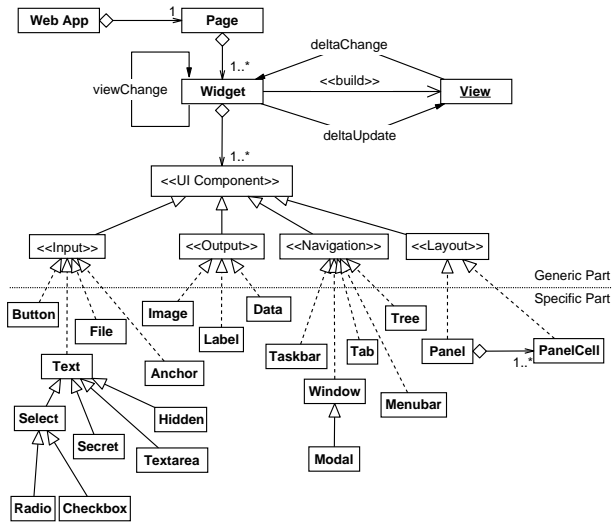

**Figure 1. The meta-model of a single-page AJAX application composed of UI components.**

2. Navigational Path Extraction
3. UI Component Model Identification
4. Single-page UI Model Definition
5. Target UI Model Transformation

Below we briefly discuss each of these steps. The main focus of this paper is on steps two and three, i.e., finding candidate user interface components to be able to define a single-page user interface model. Nevertheless, we will shortly present how we envision the other steps which are currently part of our ongoing research.

**Retrieving Pages**
Looking at dynamic web applications from an end-user's perspective enables us to gain an understanding of the application without having to cope with the many different server-side web programming languages. Building a run-time mirror-copy of the web pages can be carried out by applying static as well as dynamic analysis techniques. Static analysis can examine the pages and find `href` links to other internal pages. Dynamic analysis can help up to retrieve pages which require specific request parameters (form-based), for instance, through scenario-based test cases or collecting traces of user actions (e.g., sessions, input data) interacting with the web application. It is clear that the more our retrieved mirror-copy resembles the actual web application, the better our navigational path and UI component identification will be.

**Navigational Path Extraction**
In order to migrate from a classic web application (source) to a single-page AJAX interface (target), we first need to gain an understanding of the navigational and structural model of the source application. A *navigational path* is the route a user can take while browsing a web application, following links on the pages. For ajaxification, gaining an understanding of this navigational path is essential to be able to model the navigation in the single-page user interface model. For instance, knowing that Category pages link with Product Item List pages, implies that in our single-page model, we need a Category UI component which can navigate to the Product Item List UI component.

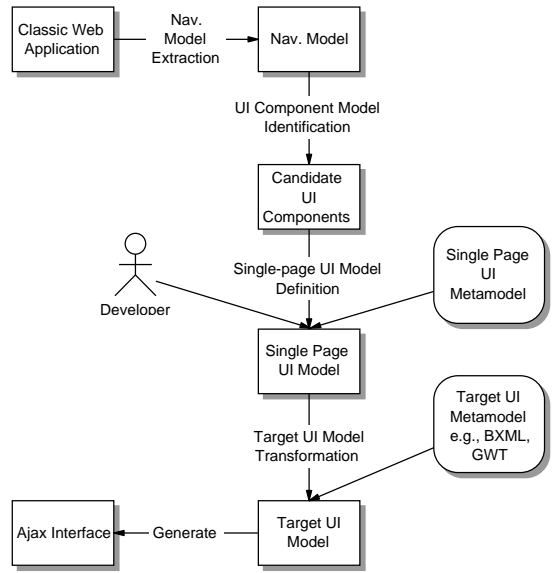

**Figure 2. Reverse Engineering Classic Web Applications to Ajax Interfaces.**

While browsing a web application, the structural changes, for certain pages, are so minor that we can instantly notice we are browsing pages belonging to a certain category e.g., Product List. Classifying these similar pages into a group, simplifies our navigational model. Our hypothesis is that such a classification also provides a better model to search for candidate user interface components.

**UI Component Model Identification**
Within web applications, navigating from one page (A) to another (B) usually means small changes in the interface. In terms of HTML source code, this means a great deal of the code in A and B is the same and only a fraction of the code is new in B. It is this new fraction that we humans distinguish as change while browsing the application.

Speaking in terms of AJAX components, this would mean that instead of going from page A to B to see the interface change, we can simply update that part of A that needs to be replaced with the new fraction from B. Thus, this *fraction* of code from page B, becomes a UI *component* on its own in our target system.

The identified list of candidate components along with the navigational model will help us define a single-page user interface model.

**Single-page UI Model Definition**
Once we have identified candidate components, we can derive an AJAX representation for them. We have opted for an intermediate single page model, from which specific Ajax implementations can be derived.

A starting point for such a model could be formed by user interface languages such as XUL[1], XIML [21], and UIML [1]. However, most such languages are designed for static user interfaces with a fixed number of UI components and are less suited for modeling dynamic interfaces as required in AJAX.

We are currently working on designing an abstract single-page user interface meta-model for AJAX applications. This abstract model should be capable of capturing dynamic changes, navigational paths as needed in such applications, and abstract general AJAX components, e.g., Button, Window, Modal, as depicted in Figure 1.

**Target UI Model Transformation**
For each target system, a meta-model has to be created and the corresponding transformation between the single-page meta-model language and the platform-specific language defined. The advantage of having an abstract user interface model is that we can transform it to different AJAX settings. We have explored [20] a number of AJAX frameworks such as Backbase[2], Echo2[3], and GWT[4], and are conducting research to adopt a model-driven approach to AJAX.

## 4. Navigational Path Extraction

Our ajaxification approach starts by reconstructing the paths that users can follow when navigating between web pages. This requires that we group pages that are sufficiently similar and directly reachable from a given page. For example, a web page *A* could contain 7 links, 3 of which are similar. We cluster those 3 pages, and look if the links contained in those 3 pages, together, could be clustered, and so on. This way we build clusters along with the navigational paths.

In this section, we discuss our specific notion of web page similarity, and the steps that we follow to compute the clusters.

### 4.1. Page Classification

Web pages can be classified in many different ways depending on the model needed for the target view. Draheim *et al.* [12] list some of possible classification notions. In this paper, our target view focuses on *structural* classification. Tonella and Ricca [23, 26] present three relevant notions of classification:

- *Textual Identity* considers two pages the same if they have exactly the same HTML code,

- *Syntactical Identity* groups pages with exactly same structure, ignoring the text between tags, according to a comparison of the syntax trees,

- *Syntactical Similarity* classifies pages with similar structure, according to a similarity metric, computed on the syntax trees of the pages.

---

[1] http://www.mozilla.org/projects/xul/
[2] http://www.backbase.com
[3] http://www.nextapp.com/platform/echo2/echo/
[4] http://code.google.com/webtoolkit/

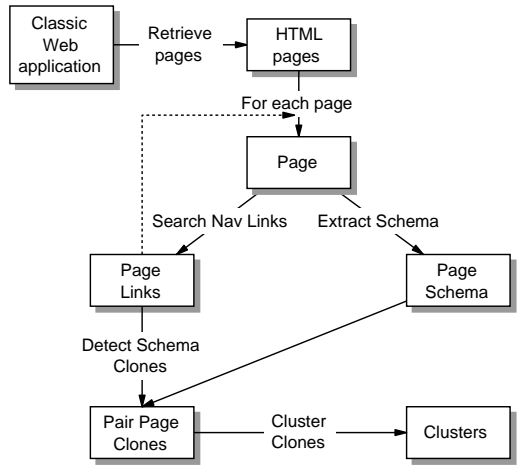

**Figure 3. Schema-based clustering process.**

Textual and Syntactical Identity classification notions have limited capabilities in finding pages that belong to a certain category as they look for exact matches. Syntactical Similarity is the notion that can help us cluster pages into useful groups by defining a similarity threshold under which two pages are considered clones. We propose a new approach based on *schema-based similarity*.

### 4.2. Schema-based Similarity

Many web clustering approaches [9, 6, 24] base their similarity degree on the computation of the edit distance between the syntax trees of web pages. This approach, although useful, has a limited capability to group HTML pages that have similar presentational structure.

For instance, consider two pages, the first page having two table rows and the second seven rows with the same structure and number of cells. On the screen, we instantly classify these two under one category, but the edit distance of these two pages could be quite high and as thus the classification metric would not classify them in one cluster. Increasing the metric threshold is not an option because that results in an increase in the number of incorrectly combined pages.

To overcome this issue, our approach relies on a comparison of the explicit schemas of pages. This means, instead of comparing the syntax trees of pages, we first reverse engineer the schemas of the pages and then compute the edit distance of the corresponding schemas. Two pages are considered clones if their schemas are similar. Going back to our example, the two pages can now be clustered correctly as a table with two row elements has the same schema as a table with seven row elements.

### 4.3. Schema-based Clustering

Given the schema-based similarity metric, we can create a schema-based clustering of web pages. We take a tree-based approach for recovering and presenting the navigational paths. In a tree structure with a set of linked nodes, each node represents a web page with zero or more child nodes and edges represent web links to other pages. We believe tree structures can provide a simple but clear and comprehensible abstract view of the navigational path for web applications.

Figure 3 shows our schema-based clustering process. Starting from a given root node (e.g., index.html), the goal is to extract the navigational path by clustering similar pages on each navigational level.

It is important to note that we do not conduct clustering of all pages at once. Instead we walk along the navigational path and for each node we cluster the pages that are linked with that node. It is important to cluster along with the navigational path, because we would like to recover the changes in the interface and later identify candidate UI components. If we cluster all pages as if they were on a single level, the navigational information will be lost and that is what we try to avoid.

Algorithm 1 shows how on each level the schemas of linked pages are compared. The search is *depth-first*. For each page on the navigational path, recursively, first the internal links (i.e., links to pages within the web application) are extracted. Afterwards, for each found link, the corresponding page is retrieved and converted to XHTML. The XHTML instance is then examined to extract an explicit schema. The variables used in the algorithm are local variables belonging to the page being processed at that level.

After the schemas are extracted, we conduct a pairwise comparison of the schemas to find similar structures. The structural edit distance between two schemas is calculated using the Levenshtein [18] method. After this step, the `connected` set contains a list of cloned pair pages (e.g., {(a-b), (b-c), (d-e)}).

To perform the classification of pages, we provide a practical way in which the actual computation of the clusters, given a set of clone pairs, i.e., `connected`, is simply done by taking the *transitive closure* of the clone relation [7]. In this approach, there is no need to define the number of clusters in advance. The result of calling the `transclos` function on our given example would be {(a-b-c), (d-e)}.

**Algorithm 1**
1: **procedure** start (Page p)
2:   Set $L \leftarrow$ extractLinks(p)
3:   **for** $i = 0$ to $L.size - 1$ **do**
4:     $pl[i] \leftarrow$ retrievePage(L(i))
5:     $px[i] \leftarrow$ convertToXHTML(pl[i])
6:     $ps[i] \leftarrow$ extractSchema(px[i])
7:     start(pl[i])
8:   **end for**
9:   Set $connected \leftarrow \emptyset$
10:  **for** $i = 0$ to $L.size - 1$ **do**
11:    **for** $j = i + 1$ to $L.size - 1$ **do**
12:      **if** distance(ps[i], ps[j]) < threshold **then**
13:        $connected \leftarrow connected \cup$ clone(pl[i], pl[j])
14:      **end if**
15:    **end for**
16:  **end for**
17:  Set $clusters \leftarrow$ transclos(connected)
18:  write(p, clusters)
19: **end procedure**

Our tool also supports an agglomerative hierarchical manner of classification. Hierarchical clustering algorithms, however, require the desired number of clusters to be defined in advance.

### 4.4. Cluster Refinement/Reduction

Because our search is depth-first, after the first page classification phase, we can further refine the clusters in order to obtain what we call the *simplified navigational path* (SNP).

Beginning at the root node, for each cluster $c$ that contains two or more pages, we examine all outgoing linked pages (from all pages in $c$) to determine whether further refinement of the classification is possible on the next levels of the navigational path. This is done by applying the same classification technique as explained in 4.3.

To simplify the navigational path, we reduce each $c$ to a node containing only one page. For that, we presently use the simple but effective approach to choose the largest page as the reduced cluster page. A more elegant (but more expensive) solution would be to replace the cluster $c$ by a page that is composed by extracting all common elements of the pages in $c$. These common elements can be computed using the shortest common supersequence algorithm [3].

From left to right, Figure 4 presents, the initial classification, the refined classification in which pages $F$, $G$, and $H$ are classified into a cluster, and the simplified navigational path (SNP) in which $Z = B \cup C$, $Y = F \cup G \cup H$, and $X = J \cup K$.

## 5. UI Component Identification

As mentioned before, our goal is to determine which parts of the web interface change as we browse from one page to another. These changes in the interface, along the navigational path, form the list of candidate components.

### 5.1. Differencing

Once a simplified model of the navigational path has been extracted, we can focus on extrapolating candidate user interface components. Using the SNP obtained in the previous step, Algorithm 2 describes how we calculate the fragment changes using a differencing approach.

**Algorithm 2**
1: **procedure** begin (Page p)
2:   $pr \leftarrow$ removeTexualContent(p)
3:   $pp \leftarrow$ prettyPrint(pr)
4:   compare(pp)
5: **end procedure**
6:
7: **procedure** compare (Page p)
8:   Set $L \leftarrow$ getLinksOnSNP(p)
9:   **for** $i = 0$ to $L.size - 1$ **do**
10:    $prl \leftarrow$ removeTexualContent(L(i))
11:    $ppl \leftarrow$ prettyPrint(prl)
12:    $candidate[i] \leftarrow$ Diff $(p \mapsto ppl)$
13:    compare(ppl)
14:  **end for**
15: **end procedure**

Starting from the root node, we compare the current page ($A$) with all the pages on the next level on the SNP to find changes between $A$ and those linked pages.

We use a modified version of the *Diff* method. This method only returns changes of going from $A$ to $B$ that are found on $B$, ignoring the changes on $A$.

To be able to conduct proper comparisons we remove all textual content, i.e., all text between the HTML tags in both pages $A$ and $B$. We also pretty print both pages by writing each opening and closing tag on a separate line.

The result is a list of candidate components in HTML code along the navigational path.

### 5.2. Identifying Elements

The list of candidate user interface components can be used, for instance, to gain a visual understanding of the changes while browsing.

The code for a candidate user interface component can also provide us useful information as what sort of HTML elements it is composed of. The elements used in the candidate components can lead us towards our choice of single-page UI components.

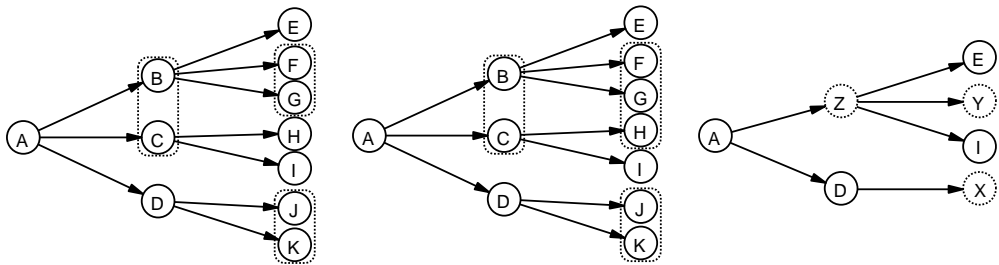

**Figure 4. Refinement and reduction of clusters.**

To that end, the content of each candidate component is examined by parsing and searching for elements of interest, (e.g., Button, Text, Textarea, Select) which can be converted to the corresponding single-page instances.

Thus, the end result of this step is a mapping between legacy HTML elements and candidate single-page user interface components.

## 6. Tool Implementation

We have implemented the navigational path extraction and AJAX component identification approach as just described in a prototype tool called RETJAX (Reverse Engineer To AJAX). RETJAX is written entirely in Java 5 and is based on a number of open-source libraries. A beta version of the tool will be made available from our software engineering site swerl.tudelft.nl.

RETJAX implements the following steps:

**Parsing & Extracting Links**
The first step consists of parsing HTML pages and extracting internal links. *HTML Parser*[5] is used for his purpose. It is also modified to pretty-print pages which is a prerequisite for the differencing step.

**Cleaning up**
For cleaning up faulty HTML pages and converting them to well-formed XHTML instances, *JTidy*[6], a Java port of the HTML syntax checker *HTML Tidy*, is used. A well-formed XHTML page is required by the schema extractor step.

**Schema Extraction**
EXTRACT [13] and DTDGenerator[7] are tools that can be used to automatically detect and generate a Document Type Definition (DTD) from a set of well-formed XML document instances. We have chosen and modified DTDGenerator to extract the schema of the XHTML version of the pages. DTDGenerator takes an XML document and infers the corresponding DTD. It creates an internal list of all the elements and attributes that appear in the page, noting how they are nested, and which elements contain character data. This list is used to generate the corresponding DTD according to some pattern matching rules.

**Distance Computation & Clustering**
The Levenshtein edit distance method is implemented in Java and used to compare the schemas pairwise. Clustering is implemented using an algorithm which finds the transitive closure of a set of clone pair.

**Simplifying Navigational Model**
After clusters have been identified, we simplify the navigational model by refining the clusters on the next levels and reducing each cluster to a single node. In the current implementation, we choose the largest page as the candidate node.

**Differencing**
The *Diff* algorithm has been implemented extending a Java version[8] of the GNU Diff algorithm. The extended version has the ability to calculate and print page specific changes between two pages. For instance, the method diff(A, B, true) returns changes in B ignoring all changes in A.

**Presentation**
The tool takes three input parameters namely, location (URI) of the initial page to start from, a similarity threshold, and the link depth level. Given these inputs, it automatically produces clusters along the extracted navigational path in DOT (Visualization) and in XML format. Also a list of connected found candidate components in XML and HTML format is produced.

## 7. Case Study

### 7.1. JPetStore

We have chosen JPetStore[9] as our migration case study, which is a publicly available dynamic web application based on Sun's original J2EE PetStore. The primary differences are that JPetStore, is vendor independent, has a standard-based multi-page web interface, and is Struts-based, which make it a typical modern web application.

---
[5] http://htmlparser.sourceforge.net/
[6] http://jtidy.sourceforge.net/
[7] http://saxon.sourceforge.net/dtdgen.html
[8] http://www.bmsi.com/java/#diff
[9] http://ibatis.apache.org/petstore.html

| Classification | # of pages |
|---|---|
| Home (Index) | 1 |
| Product Categories | 5 |
| Product Item Lists | 14 |
| Product Items | 23 |
| Checkout | 1 |
| New Account | 1 |
| Sing On | 1 |
| View Cart | 1 |
| Add Item to Cart | 24 |
| Remove Item From Cart | 24 |
| Help | 1 |

**Table 1. JPetstore Reference Page Classification.**

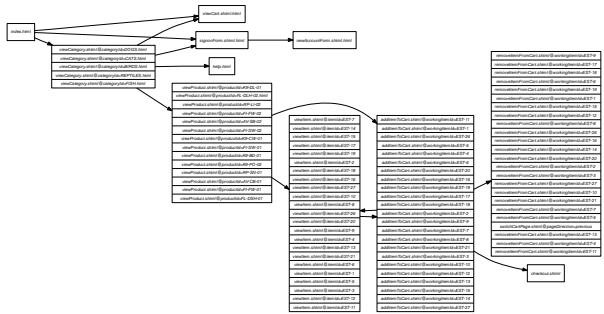

**Figure 5. Retrieved Clusters Along The Navigational Path.**

### 7.2. Reference Classification

The idea of an automatic way of supporting the migration process from multi-page to single-page web applications came to us when we initially conducted a manual re-engineering of the JPetStore web application a few months ago. Our goal was to ajaxify the application using the Backbase[10] framework.

Backbase provides a set of server-side UI components, based on the *JavaServer Faces* technology. It became immediately event to us that the fist step one needs to take in order to conduct such a migration process, is to figure out the navigational model and UI components of the current implementation of JPetStore.

Our first step was to create a mirror copy of the web application interface by retrieving as many pages as possible. A total of 96 pages were retrieved. The pages were manually examined to document a reference classification in advance. This reference classification was used for comparing candidate clusters found by the tool to evaluate the results. The reference contains 11 classifications as shown in Table 1.

---
[10] http://www.backbase.com

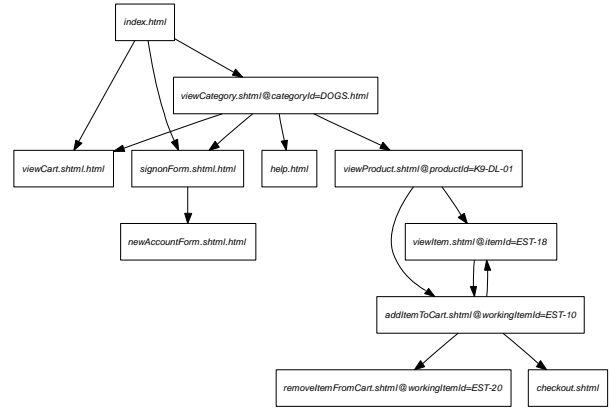

**Figure 6. Reduced Clusters.**

### 7.3. Automatic Classification

The aim of the case study is to determine to what extent we can use JPetStore's web interface to automatically find a list of candidate user interface components along with their navigational path.

In order to conduct a preliminary evaluation of the described reverse engineering process, we used two different methods, namely, our own schema-based similarity approach (MMS), and our own implementation of a syntax tree similarity (STS) approach as proposed by, e.g., [8]. We also used different thresholds to find out the best achievable results.

In the first step of the reverse engineering process, pages were clustered along the navigational path (tree-based) and the navigational path was reduced by refining the clusters, as shown in Figure 5. Subsequently, as illustrated in Figure 6, in the second step, found pages in each cluster were reduced to a single node using the method described in 4.4.

Afterwards, candidate UI components were calculated by applying the *Differencing* algorithm as described in Section 5.

Figure 7 depicts viewing a candidate UI component (HTML code) in a browser, which is the result of going from the index page to the (dogs) category page. As a result, only that fraction of the category page that is unique with respect to the index page is reported. This way, we are able to visualize the delta changes (candidate single-page components) of the web interface by browsing the navigational path.

The list of candidate components and the simplified navigational path help us model our target single-page interface in the Conallen UML extension [4], which is shown in Figure 8. Our single-page (called Page), contains three UI components namely, Category, Cart, and SignOn. Navigation takes place by changing the view from one component to another. For instance, from the Category component we can change our view to go to the Product component. This is a delta change, meaning only that part of the Page that con-tained the Category component will be updated to view the new Product component.

**Figure 7. A candidate UI component (Product Category).**

**Figure 8. Target JPetstore Single-page Interface.**

### 7.4. Evaluation

Two well known metrics namely *precision* and *recall* were used to evaluate the results. Precision represents how accurately the clusters from the algorithm represent the reference classification. Recall measures how many pages in the reference classification are covered by clusters from the algorithm. We count only exact matches against the reference classification in the *Relevant Clusters Retrieved* (RCR) group. This means, if the algorithm finds a cluster which contains one or more extra (or one or more less) pages than the corresponding reference cluster, it is counted in the *Irrelevant Clusters Retrieved* (ICR).

Other comparison techniques, such as the ones introduced by Koschke and Eisenbarth [16] and Tzerpos and Holt [27] could also have been chosen. However, we would expect similar results from these techniques as well.

Table 2 shows the results. With the syntax tree similarity (STS) approach, the best recall value obtained was 82 % with a precision of 69 %, using a similarity threshold of 91 %.

The meta-based similarity (MMS) approach, however, was able to find all 11 documented reference clusters with a recall and precision of 100 % using a similarity threshold of 98 %. Note that by increasing the threshold to 99 %, the precision and recall drop to 82 %. This behavior can be explained because the algorithm expects the schemas to be almost identical, and as a result very little difference in the corresponding pages is tolerated. This increases the number of false positives.

## 8. Discussion

As mentioned before, we came to the idea of a tool support for ajaxification process when we first conducted a manual migration.

The required knowledge for ajaxification was obtained by manually browsing the interface, from one page to the other, noting the differences, and building a map of the interaction model. This was when we realized that reverse engineering techniques should be able to provide some degree of support. Having a tool which provides us with information about the UI components needed and their positions on the navigational paths, can be of great value.

Applying the techniques described in this paper to our case study, we were able to find all reference classifications. Additionally, with some degree of manual intervention, we were able to create a single-page model of the target system.

Even though the techniques introduced in this paper have only been applied to one case study, considering the results obtained, we believe the applications can span real-world web application migration cases. Although the JPetStore interface is very simple, it is representative of dynamic transactional web applications, and this class of web applications is exactly what we aim for. Our approach is not meant for web sites that are composed of long pages such as news, article, or forum sites. We will need to conduct more case studies to find strengths and weaknesses of our techniques and improve the tool.

We take a client-side analysis approach. While having the benefit of being server-code independent, the information that can be inferred from the server-side, such as scripting languages as JSP, is also essential for conducting a real migration process.

One of the problems we encountered while carrying out the case study, was that some HTML pages contained elements that were not well-formed or were not recognized by the formatter. Even JTidy was not able to fix the problems and no conversion to XHTML could be conducted. For instance in a few pages, instead of `` element a `<image ...>` was used. Manual intervention was required to fix the problem. This sort of problems are inherent in web applications and can cause real problems in real-world migration cases, where standard guidelines are neglected and faulty HTML code is written/generated.

| Method | Threshold | RCR | ICR | Precision (%) | Recall (%) |
|---|---|---|---|---|---|
| STS | 0.89 | 6 | 3 | 66 | 54 |
| STS | 0.91 | 9 | 4 | 69 | 82 |
| STS | 0.93 | 7 | 8 | 46 | 63 |
| MMS | 0.97 | 7 | 1 | 87 | 63 |
| MMS | 0.98 | 11 | 0 | 100 | 100 |
| MMS | 0.99 | 9 | 2 | 82 | 82 |

**Table 2. Results of Clustering JPetstore Web Interface.**

## 9. Related Work

Reverse engineering techniques have been applied to web application settings primarily to gain a comprehensible view of the systems.

Hassan and Holt [15] present an approach to recover the architectural model of a web application by extracting relations between the various components and visualizing those relations.

Di Lucca *et al.* [10, 11] propose WARE which is a tool for reverse engineering Web applications to the Conallen extension [4] of UML models. Draheim *et al.* [12], present Revengie to reconstruct form-oriented analysis models for web applications.

Ricca and Tonella [23] propose ReWeb, a tool to analyze source code to recover a navigational model of a web site. They use the models obtained by ReWeb for testing [26] web applications. Supporting the migration of static to dynamic web pages is illustrated in [24] by applying an agglomerative hierarchical clustering approach.

De Lucia *et al.* [7, 8] present a program comprehension approach to identify duplicated HTML and JSP pages based on a similarity threshold using Levenshtein string edit distance method. They use three notions of similarity namely, structure, content, and scripting code. In [6], they apply the techniques in a re-engineering case study.

WANDA [2] is a tool for dynamic analysis of web applications. It instruments web pages and collects information during the execution. This information is used to extract diagrams, such as component, deployment, sequence and class diagrams according to Conallen UML extensions.

Cordy *et al.* [5] use an island grammar to identify syntactic constructs in HTML pages. The extracted constructs are then pretty-printed to isolate potential differences between clones to as few lines as possible and compared to find candidate clones using the UNIX diff tool.

A study of cloning in 17 web applications is presented by Rajapakse and Jarzabek [22], aiming at understanding the nature of web clones and their sources. Lanubile and Mallardo [17] discuss a pattern matching algorithm to compare scripting code fragments in HTML pages.

Stroulia *et al.* [25] analyze traces of the system-user interaction to model the behavior of the user interface for migrating the user interface from a legacy application to a web-based one. GUI Ripping [19] creates a model from a graphical user interface for testing purposes, i.e., it generates test cases to detect abnormalities in user interfaces. Vanderdonckt *et al.* [28] propose Vaquista, a XIML-based tool for static analysis of HTML pages. Its goal is to reverse engineer the user interface model from individual HTML pages to make them device independent.

Our classification approach is in two ways different from work conducted earlier on this topic. First, while others have based their structural similarity notion on the edit distance calculated on the syntax trees of pages, we propose a meta-model similarity notion and implement a schema-based clustering approach which, in the case of HTML pages, provides very promising results. Second, we try to find the clusters along the navigational path (different levels), as opposed to classifying all pages at once (one level) in order to identify candidate UI components along with their navigational model.

## 10. Concluding Remarks

In this paper, we emphasized the rise of single-page AJAX applications and the need for support in migrating classical multi-page web applications to this new paradigm.

**Contributions**

The main contributions of this paper can be summarized as follows. First, we proposed a migration process, consisting of five steps: retrieving pages, navigational path extraction, user interface component model identification, single-page user interface model definition, and target model transformation. The second and third steps were described in full detail.

Second, we introduced a novel meta-model similarity metric for web page classification, which in our case studies achieves a higher recall and precision than approaches based directly on the HTML syntax trees.

Third, we proposed a schema-based clustering technique that operates per navigational level, instead of on the full set of web pages. Furthermore, we provide a mechanism for simplifying navigational paths, allowing us to find candi-date user interface components through a differencing mechanism.

Our fourth contribution is the RETJAX tool that implements the approach. Last but not least, we have used RETJAX to apply our approach to the JPetStore case study. With an appropriate threshold, we were able to fully reconstruct a reference classification automatically.

**Future Work**
Future work encompasses the in-depth application of our approach in other case studies. We will focus on the last two steps of the proposed migration process and study how a model-driven approach can be adopted in AJAX development. We are in the process of making our tool publicly available on the Web. Furthermore, we will investigate how we can take advantage of dynamic analysis concepts to support the retrieving pages step of the migration process.

Finally, we will conduct research as how and to what extent the server-side code should be adapted while migrating from a multi-page web application to a single-page AJAX interface.

**Acknowledgments** *This work received partial support from SenterNovem, project Single Page Computer Interaction (SPCI). We thank the anonymous referees for their valuable comments.*

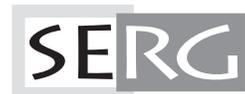